# The AOLI low-order non-linear curvature wavefront sensor: a method for high sensitivity wavefront reconstruction

Jonathan Crass[*a], Peter Aisher[a], Bruno Femenia[b,c], David L. King[a], Craig D. Mackay[a], Rafael Rebolo-López[b,d], Lucas Labadie[e], Antonio Pérez Garrido[c], Marc Balcells[f,b,g], Anastasio Díaz Sánchez[c], Jesús Jimenez Fuensalida[b,g], Roberto L. Lopez[b], Alejandro Oscoz[b,g], Jorge A. Pérez Prieto[b,g], Luis F. Rodríguez-Ramos[b], Isidro Villó[c].

[a]Institute of Astronomy, University of Cambridge, Madingley Road, Cambridge, CB3 0HA, UK.
[b]Instituto de Astrofísica de Canarias, C/ Via Lactea s/n, La Laguna, Tenerife E-38200, Spain.
[c]Univ. Politécnica de Cartagena, Campus Muralla del Mar, Cartagena, Murcia E-30202, Spain.
[d]Consejo Superior de Investigaciones Científicas, Spain.
[e]I. Physikalsiches Institut, Universität zu Köln, Zülpicher Strasse 77, 50937 Köln, Germany.
[f]Isaac Newton Group of Telescopes, Apartado de Correos 321, Santa Cruz de la Palma, Canary Islands E-38700, Spain.
[g]Univ. de La Laguna, Pabellón de Gobierno, C/ Molinos de Agua, La Laguna E-38200, Spain.

**ABSTRACT**

The Adaptive Optics Lucky Imager (AOLI) is a new instrument under development to demonstrate near diffraction limited imaging in the visible on large ground-based telescopes. We present the adaptive optics system being designed for the instrument comprising a large stroke deformable mirror, fixed component non-linear curvature wavefront sensor and photon-counting EMCCD detectors. We describe the optical design of the wavefront sensor where two photon-counting CCDs provide a total of four reference images. Simulations of the optical characteristics of the system are discussed, with their relevance to low and high order AO systems. The development and optimisation of high-speed wavefront reconstruction algorithms are presented. Finally we discuss the results of simulations to demonstrate the sensitivity of the system.

## 1. INTRODUCTION

Adaptive optics (AO) systems are now commonplace on large ground-based telescopes providing correction for the effects of atmospheric turbulence. These systems comprise three main components – a sensor to take measurements of the incoming wavefront from a reference object, a control system to process the measurement data and determine the distortion of the wavefront and finally a corrector to apply the determined correction onto the wavefront.

The majority of current AO systems employ one or more Shack-Hartmann wavefront sensors (SHWFS) to determine the atmospheric distortions on incident wavefronts. These sensors typically require a bright reference star (I~12-14 magnitude). The scarcity of natural guide stars at this magnitude or brighter has led to the development of laser guide stars [1]. Although laser guide stars have been used with some success, particularly at near infrared wavelengths, there are still difficulties to overcome (*e.g.* the finite size of the laser beacon, focal anisoplanatism).

*jcrass@ast.cam.ac.uk

Guyon [2] simulated alternative wavefront sensing methods that offer substantial benefits in sensitivity when compared to a traditional SHWFS. Most significant was the non-linear curvature wavefront sensor (nlCWFS) requiring ≈100-1000 times fewer photons than SHWFSs when correcting for low-order aberrations on 8-10m class telescopes (see Section 2). Racine [3] has also shown significant improvements in sensitivity for deployed curvature wavefront sensors compared to SHWFSs, particularly at low orders. These sensitivity improvements increase the limiting magnitude for any reference object and lead to a significant increase in the number of natural guide stars available as reference objects, eliminating the need for laser guide stars. However, even with this improvement it is still challenging to significantly improve the image quality in the visible using AO alone as a residual RMS error of ~100nm is required.

An alternative method for correcting atmospheric effects on astronomical images is the technique of Lucky Imaging. Turbulence in the atmosphere is random in nature and hence probabilistic. By taking a series of short exposure images (~30Hz), some of these images will exhibit reduced turbulent effects. Images may be ranked using a 'sharpness characteristic' (*e.g.* the FWHM of a reference object) and then a percentage of the best images are combined to give a sharper output image.

Lucky Imaging works well on telescopes up to ~2.5 metres in diameter. With larger diameters, the increasing number of turbulent cells across the telescope reduces the likelihood of obtaining an image with minimal atmospheric effects. This makes the technique impractical on larger telescopes due to the increased observing time required. In addition, one of the factors limiting the effectiveness of Lucky Imaging is that a selection criterion has to be used meaning a large amount of data is discarded, reducing the sensitivity of observations. Garrel, Guyon and Baudoz [4] have recently suggested that image selection may be more effective in Fourier space allowing the retention of a greater percentage of the data, and improving the sensitivity. This has been verified through the reprocessing of existing Lucky Imaging data [5].

It is useful to consider the probability of obtaining a 'lucky exposure' where the RMS phase error across a telescope aperture of *D* metres is less than 1 radian. The probability was defined by Fried as

$$p \simeq 5.6 \, exp\left[-0.1557 \left(\frac{D}{r_0}\right)^2\right] \quad (1)$$

which for a 2.5m telescope, assuming $r_0$ = 35cm, gives $p$ = 0.002 [6]. For Lucky Imaging to be successful on larger telescopes this probability needs to be maintained and as such, the atmospheric effects need to be reduced to be of the order of those on a 2.5m telescope. This means reducing the effective number of turbulent cells over the telescope aperture. This can be achieved by placing a Lucky Imaging detector behind a low order AO system, a technique previously demonstrated using PALMAO and LuckyCam on the 5m Hale telescope [7] and with NAOMI and Fastcam on the 4.2m William Herschel Telescope (WHT) [8]. Simulations have also shown the combination of AO and Lucky Imaging techniques on 10m class telescopes gives a gain in Strehl Ratio. The relative gain is largest at shorter optical wavelengths, doubling the Strehl Ratio in the R-band to 12% and delivering a value of over 20% at I-band when compared to AO alone [9].

The Adaptive Optics Lucky Imager (AOLI) instrument is designed to combine the techniques of AO and Lucky Imaging into a single instrument. The AO component of the system employs a nlCWFS comprising pairs of beam splitting optics, two Electron Multiplying CCDs (EMCCDs) to maximise its sensitivity to natural guide stars and an ALPAO deformable mirror to apply wavefront corrections. The instrument development is a collaboration between several European research institutions: Instituto de Astrofisica de Canarias/Universidad de La Laguna (Tenerife, Spain), the Universidad Politecnica de Cartagena (Spain), Universität zu Köln (Germany), the Isaac Newton Group of Telescopes (La Palma, Spain) and the Institute of Astronomy, University of Cambridge (UK). It is initially designed specifically for use on the WHT and the 10.4m Gran Telescopio Canarias (GTC). It is estimated that 17.5-18th magnitude natural guide stars will be usable as reference objects on the WHT with the limiting magnitude for the GTC being around 1 magnitude fainter.

This paper describes the adaptive optics component of the instrument and in particular the nlCWFS (Section 2 and 3), its optical setup (Section 4) and algorithms for determining the wavefront distortions from the EMCCD measurements (Section 5).

## 2. NON-LINEAR CURVATURE WAVEFRONT SENSING

Most adaptive optics systems today are designed to give a high degree of correction to incoming distorted wavefronts. Most use a SHWFS which comprises an array of subapertures to break up the incident light into a number of cells, each of which is imaged onto a CCD. By locating the centroid of the reference object on each region of the CCD, the localised distortion of the wavefront can be calculated. However, as the light is shared between hundreds or thousands of cells, the intensity per cell is reduced.

The curvature wavefront sensor (CWFS) was first proposed by Roddier [10]. In this method, an image is taken on either side of a focal plane of a reference object. By comparing the changes in light intensity on either side of the focal plane, the curvature of the incoming wavefront may be derived.

Guyon proposed a different method of determining wavefront distortions using a non-linear curvature wavefront sensor (nlCWFS) [11]. Here, a total of four images are taken; two on either side of a conjugate pupil plane (see Figure 1). The use of the four planes allows both low and high order aberrations to be determined with improved sensitivity compared to both SWHFS and more conventional CWFS methods (see Figure 2).

The nlCWFS uses a phase retrieval algorithm to determine the phase at the pupil plane from intensity measurements. An initial wavefront estimate is repeatedly propagated (in the Fresnel regime) between the four image planes and the pupil plane with constraints in intensity or phase being applied after each propagation. The procedure is performed iteratively, converging upon the actual wavefront distortion.

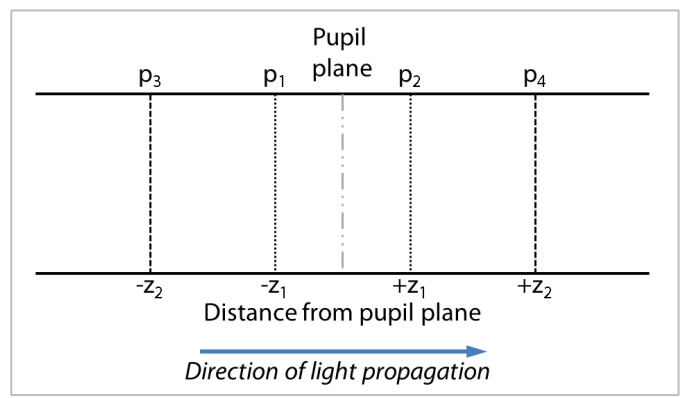

Figure 1: Location of the image planes ($p_1$-$p_4$) relative to the pupil. The image planes are located in an inner and outer pair with the distance for each image in the pair being equidistant from the pupil.

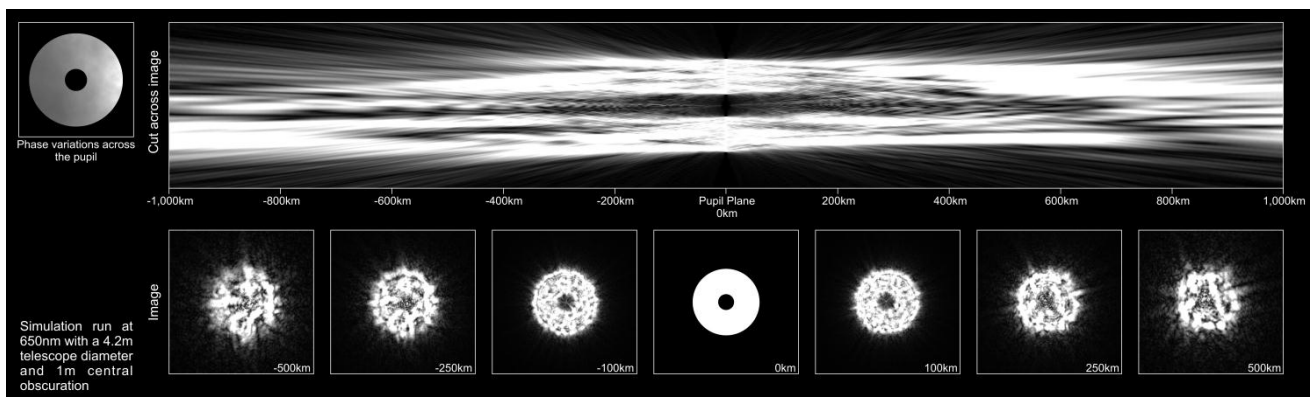

Figure 2: The propagation of light (intensity) from the pupil plane of a telescope. The pupil (shown in the centre) has uniform illumination and a phase value to represent the propagation through a Kolmogorov simulated turbulent atmosphere. On either side of the pupil, the intensity breaks into speckles. The smaller speckles show the higher order structure and are located nearer to the pupil while the larger structure, which shows the low orders, develops as the propagation distance is increased.

The first algorithm to retrieve phase from intensity measurements was the Error-Reduction (ER) algorithm developed by Gerchberg and Saxton [12]. Although successful, the algorithm can stop converging for several iterations, slowing down reconstruction, a particular problem for the correction of atmospheric effects in real time. Further algorithms for phase retrieval from intensity have been proposed to counter this problem. They are all principally based on the ER algorithm with modifications to the iterative component [13].

Initial work by Guyon used the standard ER algorithm for phase reconstruction [11]. This was successful using monochromatic light and polychromatic light when corrective optics were applied. Corrective optics are required when determining higher order aberrations as the inherent chromatic effects in Fresnel propagation cause the image speckles to appear blurred. As part of the development of the AOLI, further simulations have been undertaken to characterise the nlCWFS method of phase retrieval and investigate chromatic effects (Section 3).

Most types of wavefront sensor work well with many incident photons. However, at low photon counts many of them fail including the nlCWFS. This failure arises due to phase reconstruction algorithms performing poorly with low photon numbers as zero intensity values are imposed as a constraint during each iteration. This leads to a loss in phase information. Techniques to overcome this problem have been investigated and are discussed in Section 5.

## 3. WAVEFRONT SENSOR SIMULATIONS

Understanding chromatic effects is important when selecting the appropriate bandpass for a reference object where sensitivity is a key consideration. These effects were investigated by simulating the image plane locations for the nlCWFS using the Arroyo simulation package [14] and the following procedure:

1. Generate a Kolmogorov model turbulent phase screen applying sub-harmonic correction as proposed by Lane *et al* [15] to simulate the turbulent atmosphere
2. Pass a monochromatic plane wave through the generated phase screen
3. Apply an aperture mask to the wave, representing the aperture of the telescope
4. Propagate to each plane in turn using Fresnel propagation methods.

Monochromatic simulations were performed using light with wavelengths of 500, 700 and 900nm with the same turbulent phase screen at the pupil. The wavefront was propagated to multiple locations either side of the pupil where the amplitude was recorded. Preliminary results are shown in Figure 3.

The simulations show a divergence in the intensity of different wavelengths as the propagation distance is increased. Up to around 100km, the effects are limited to a blurring effect. However, as distances reach that of the non-linear regime, the effects become more pronounced with a significant difference between wavelengths.

These effects, while initially concerning, may not be as problematic as they seem. Corrective optics can be used to remove the effects and this is only required if performing high order correction [16]. If only low-order correction is required, as seen in Figure 3, the fundamental low-order intensity structure is maintained even at large distances. As such, it appears corrective optics may not be required in the low-order regime.

As the AOLI instrument uses a combination of wavefront correction techniques, the AO component is primarily focussed on performing low-order correction. Because of this, we intend as part of the reconstruction algorithm development to investigate whether low-order correction is still possible without the need for corrective optics. In addition, rather than using conventional beamsplitters to provide the four image planes (Section 4), it may be beneficial to split in wavelength range rather than amplitude. However, this will likely add additional computational complexity when reconstructing.

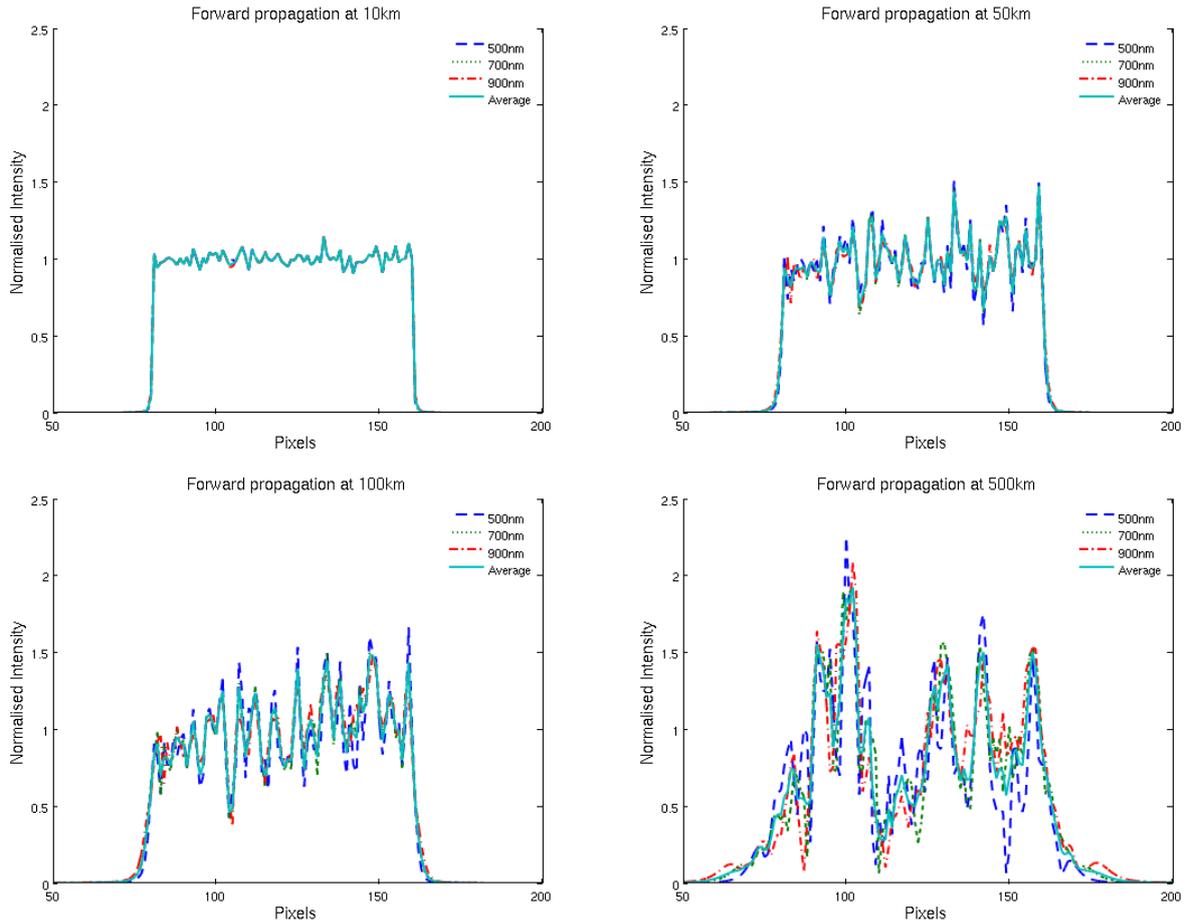

Figure 3: Simulation showing the intensity difference between 500, 700 and 900nm wavelength propagations based on the same pupil wavefront. The figure shows the horizontal cut across the images at 10, 50, 100 and 500km for an 8m telescope with uniform illumination and a turbulent phase. Close to the telescope pupil, there are only minor chromatic effects however at larger distances, the higher order structure becomes smeared out. The low-order structure is still visible however.

## 4. THE AOLI WAVEFRONT SENSOR – OPTICAL SETUP

The current optical design for AOLI for use on the WHT is shown in Figure 4. The incident light from the telescope initially passes to a 97 element (11 elements across the diameter) ALPAO deformable mirror (DM97-15). This mirror has been selected due to its large stroke (±60 micron tip/tilt correction), alleviating the need for a tweeter and woofer configuration [17].

A pickoff mechanism is used to divert light from the reference object (located on the optical axis of the telescope) to the wavefront sensor. Depending upon the magnitude of the reference object and seeing conditions, different spot sizes and optical densities of the pickoff device can be used. This allows some light to pass to the Lucky Imaging detector from the reference object when possible.

The wavefront sensor assembly is shown in Figure 5. A cube beamsplitter diverts the light to two subsequent beamsplitters, each generating a pair of image planes. The pentaprism beamsplitter generates the outer pair of images ($p_3$ and $p_4$ in Figure 1) while a lateral displacement beamsplitter produces the inner pair ($p_1$ and $p_2$).

The output from each beamsplitter is imaged onto a single back-illuminated 1024×1024 pixel EMCCD manufactured by E2V Technologies Ltd (Model CCD201). Readout is done using custom electronics developed for the LuckyCam instrument in Cambridge allowing up to a 30MHz pixel rate and a full frame rate of 25 frames per second. The size of each image pair on the EMCCD will be much smaller than the overall imaging region of the device and by using a limited readout region, frame rates of greater than 100Hz can be achieved.

Although the setup presented is designed for use on the WHT, the layout for use on the larger 10.4m GTC requires minimal design changes. As the final residual error from the AO system needs to be maintained at the same level as on the WHT to allow for Lucky Imaging, each actuator mapping across the telescope pupil must be approximately the same scale. As such, it is proposed to use a 277 element (19 elements across the diameter) DM277-15 ALPAO deformable mirror to match this requirement for the 10.4m aperture.

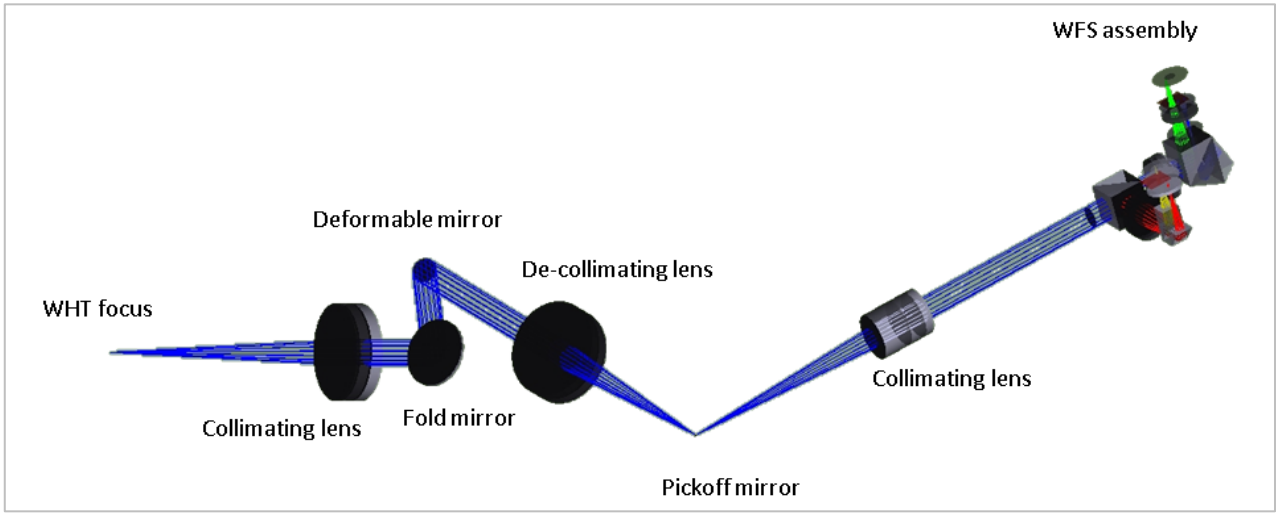

Figure 4: The full adaptive optics system for AOLI. Incoming light from the WHT is incident onto a 97 element DM from ALPAO. The reflected wavefront is reimaged to a focal point where a pickoff mirror mechanism is used to redirect the reference object light to the nlCWFS setup. The remaining light passes to the Lucky Imaging science camera (not shown).

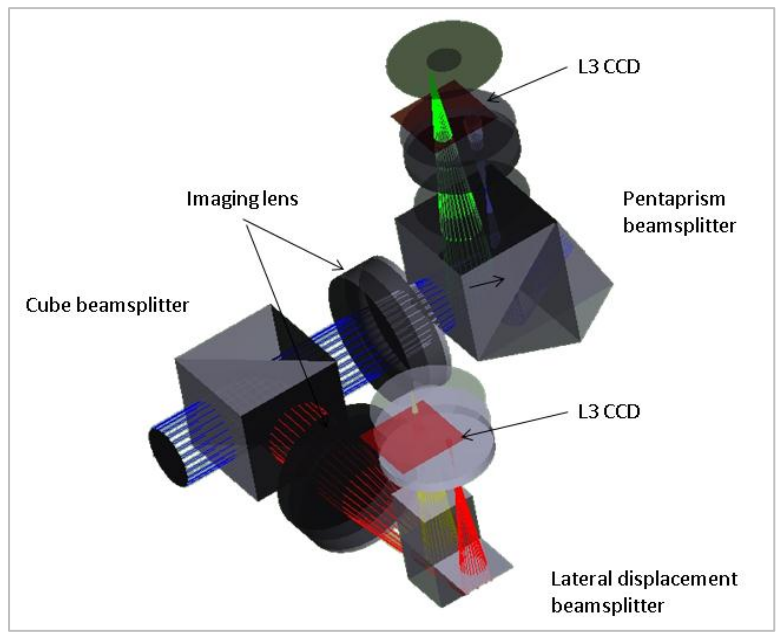

Figure 5: The wavefront sensor optics showing both sets of beamsplitters providing the four image planes. A pentaprism beamsplitter provides the outer pair of images while a lateral displacement beamsplitter provides the inner.

# 5. RECONSTRUCTION ALGORITHMS

Previous reconstruction methods for the nlCWFS have used a four plane ER algorithm operating at speeds of the order 1Hz [18]. For use on-sky, the process must converge rapidly for low order correction and with this requirement, significant work has been focussed on the development of a fast algorithm for reconstruction.

Fienup [13, 19] suggested improved algorithms for phase reconstruction overcoming the principal drawback of stagnation in the ER algorithm, namely the Input-Output (IO) and Hybrid-Input-Output (HIO) algorithms. Instead of simply imposing the known constraints *i.e.* the intensity and aperture mask, these methods compare the difference between the input and output prior to propagation, constraining and inverse propagation. This technique significantly improves the convergence speed and has been the basis of further algorithm development work. Investigations have focussed on both the reconstruction speed and appropriate techniques for low photon counts.

## 5.1 Reconstruction Speed

The method for wavefront reconstruction is shown in Figure 6. Each plane is visited systematically with constraints being applied at each location. A new estimate for the wavefront is generated after each stage and varies depending upon which algorithm is used *i.e.* ER, IO, HIO *etc*. Unless stated otherwise, all tests have been completed using the Input-Output algorithm.

The principal computational cost is the propagation between planes since for each propagation two Fast Fourier Transforms are required. As such, minimising the number of iterations required is vital for a fast-running algorithm. Benchmark tests have been undertaken comparing the order in which the planes were visited, offering little change in convergence speed. However, by applying the aperture mask as a constraint on phase and intensity whenever a propagation passed through the pupil, the rate of convergence increased significantly.

A further development of the Input-Output algorithm has been proposed where additional information from previous iterations of the reconstruction process is used. This 'Input-Output-Output' algorithm has shown a further increase in the rate of convergence, particularly at low photon numbers [20].

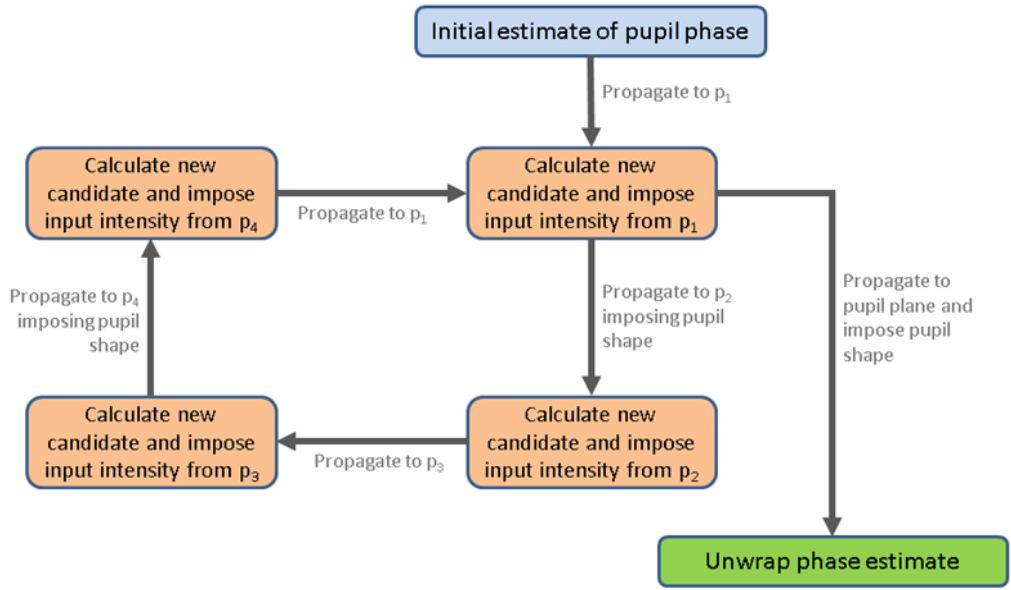

Figure 6: A schematic diagram of the phase reconstruction algorithm for four planes as described in Figure 1. The new candidate generation after each propagation depends upon the specific algorithm being used. It was found that by applying constraints whenever a propagation passed through the pupil led to a significant increase in convergence speed.

## 5.2 Low Photon Number Reconstruction

In the regime of low photon rates, binning of pixels on CCD detectors is typically used. By doing this however, the effective pixel spacing of an image is increased and above a certain level this increases errors in phase retrieval algorithms. As such, investigations have focussed on using pre-processing techniques with unbinned data prior to the use of a phase reconstruction algorithm.

Several processing methods are available to interpolate image data with small numbers of detected photon events. With the AOLI wavefront sensor using EMCCDs in photon-counting mode, it is possible to use a Gaussian convolution (GC) for each photon detected. In addition, as the image plane pairs are at different distances from the pupil, each will have different speckle scales, so instead of a fixed convolution scale being used, different convolution scales can be used for each pair *i.e.* differential Gaussian convolution (DGC). Further methods of interpolation include area-weighted triangulation filling (AWTF) using Delaunay triangulation and polygon filling using Veronoi tessellation (AWVF).

We have investigated these pre-processing methods using the Input-Output algorithm in the standard reconstruction process. The DGC and AWTF algorithms were applied to simulated low photon number data before being used in the reconstruction. Both methods were compared to the standard method without pre-processing.

Both methods show significant improvements over the standard Input-Output algorithm as shown in Figure 7. DGC can be thought of as a 'smearing' effect preserving the location of the photon count while AWTF (and AWVF) distribute the photon intensity over a region. Both methods are particularly effective for low-order correction as these modes are more spatially spread over the image plane. AWTF offers a slight benefit compared to DGC although due to the processing involved it requires significantly more computing time.

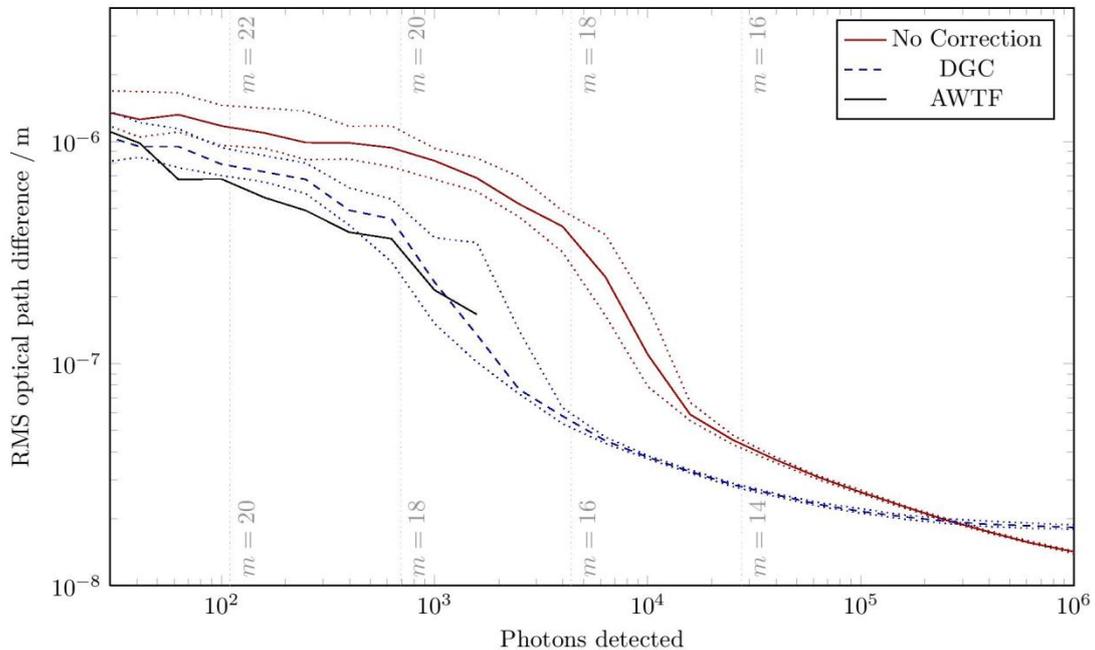

Figure 7: Simulated low photon number performance of the Input-Output algorithm with and without pre-processing. DGC, AWTF and no pre-processing provide little or no correction up until around 100 photons are detected. Beyond this, the pre-processing methods begin to perform better until the high photon regime where the conventional method has a lower RMS. The dotted lines show the upper and lower quartiles. The AWTF method was only used below 2000 photons. Vertical lines show the I-band magnitude for a CWFS running at 10Hz for D=4.2m (lower label) and D = 10.5m (upper label).

At high photon rates per frame, the additional pre-processing limits phase convergence and the standard Input-Output algorithm is more effective. Therefore, any control system must be able to turn these algorithms on as required to match the reference object and seeing conditions. Even in the low photon count regime, there are specific ranges where certain algorithms were better suited, so it appears unlikely there is a 'one algorithm fits all' solution. A range of algorithms and parameters which can be activated as required will allow the optimum reconstruction.

### 5.3 Further algorithm development

All algorithm work to date has been undertaken with no prior knowledge of the likely shape of the wavefront in the pupil plane. When observing on-sky, the overall change between frames is likely to be small. It is expected that once a reasonable correction has been achieved, convergence on the next phase estimate should occur in fewer iterations by using the previous correction as prior knowledge. The remaining challenge is to reach the initial stage of having a reasonable estimate. Further investigation of this is planned.

A statistical reconstruction (e.g. Bayesian) algorithm may further improve the results of reconstruction particularly in the low photon count regime. In addition, it may be possible to use parallel processing methods to perform all propagations and constraints simultaneously before averaging to provide the next best estimate. These methods are yet to be investigated.

Although the algorithm used for reconstruction is important, one of the principle reasons for fast phase reconstruction is the use of the four planes. This provides more information to constrain the problem, a key factor in improving the speed.

## 6. CONCLUSIONS

The combination of the corrective techniques of adaptive optics and Lucky Imaging have previously been shown to offer substantive benefits in improving resolution of ground-based telescopes at optical wavelengths. We believe that the AOLI instrument, which combines the two techniques in a single instrument for the first time, will demonstrate near-diffraction limited imaging in the visible on large ground-based telescopes. The use of a new non-linear curvature wavefront sensor in conjunction with new selection techniques for Lucky Imaging will help further improve resolution and by running the systems together, information regarding the quality of the wavefront could be used in the Lucky Imaging selection process. We estimate the system will be able to use natural guide stars as faint as 18.5-19$^{th}$ magnitude and provide an image resolution of ~0.015 arcsec on the 10.4 m Gran Telescopio Canarias at optical wavelengths.

The success of the wavefront sensor in this regime requires the development of algorithms to work in the wide band low photon regime using EMCCD detectors. Work to date has shown the use of pre-processing methods before phase-reconstruction offers benefits over conventional CCD binning. Furthermore, the use of statistical methods would further aid in improving the reliability and speed of reconstruction. Further work is being undertaken to develop these methods through simulations using simulated on-sky data and a laboratory demonstrator introducing known wavefront distortions.